\documentstyle[12pt]{article}
\newcommand{\vek}[1]{\mbox{\boldmath $   #1$}}
\newcommand{\etal}{{\it et al\/}\ }
\newcommand{\dash}{------}
\newcommand{\e}{{\mathrm e}}

\begin{document}  

\title{{\bf The energy dependence of relativistic  nonradiative
electron capture}}

\author{
J F McCann \dag,
J T Glass\ddag and
D S F  Crothers \ddag \\
\dag Department of Physics, 
University of Durham,  \\
Durham DH1 3LE, England. \\
 \ddag 
The Theoretical and Computational Physics Research Division,  \\
The Queen's University of Belfast,\\
 Belfast BT7 1NN, Northern Ireland.}

\date{April 1996}

\maketitle

\begin{abstract}
The energy dependence of the nonradiative electron capture 
cross-section is discussed in the 
relativistic domain.
A simple analytic expression is obtained  
for inner-shell transitions using second-order perturbation theory. We 
have confirmed that the leading-order term is found
to have the following energy dependence: $\sigma \sim E^{-1}  \ln^2 E$.
This is attributed to a combination of 
kinematic features of the process and  
retardation effects. 
Electron capture without change of spin is the dominant transition.
\end{abstract}

\noindent
PACS~~31.15.Md, 31.30.Jv, 34.70+e, 52.20.Hv

%\newpage
The asymptotic form of the
electron capture process for relativistic energies 
has been a subject of debate for some time. An overview 
of this question has been given by Mukherjee \etal (1992) and
Bransden and McDowell (1993), while a detailed discussion
is given by Eichler and Meyerhof (1995). 
In the relativistic domain, the physical constraints of 
energy and momentum matching conspire to make capture a 
weakly-coupled process in comparison to ionization or excitation.
Indeed it is interesting to remark that
for relativistic energies, and for small 
nuclear charges, vacuum coupling in the form of radiative electron capture 
is the dominant mechanism for charge exchange.
As a consequence, the numerical results are primarily of academic interest in
the extreme relativistic domain. 
Nonetheless, results for the nonradiative process 
ultimately provide important benchmarks and
calibration standards for both experiment and
theory (Eichler and Meyerhof 1995). 
Moreover the phenomenon  has attracted  attention in 
recent years not least because it is physically important for
large nuclear charges, and this makes it 
accesssible to experimental studies up to and beyond the 
energy range GeV/u (Anholt 1985). 
In the following
discussion we use the Lorentz factor $\gamma$ as the energy 
parameter, and this is related to the laboratory frame 
kinetic energy through the usual relation
 $E=(\gamma-1)M_P c^2$, with $M_P$ the projectile mass.  

Experimental data for relativistic collisions 
have been analysed and compared with theoretical models (Eichler 1990,
Moiseiwitsch 1989, Deco and Rivarola 1987, Glass \etal 1994),
and satisfactory
agreement has been obtained in many cases. 
However an intriguing question still remains open regarding 
the high-energy limit of the cross section, both in terms of 
the energy dependence and the size of the cross section.
The uncertainty over this result 
reflects a  well-known anomaly for high-energy nonrelativistic 
collisions, namely that  
electron capture is dominated by second-order terms. 
Nonetheless the dominance of the second-order term, 
via the Thomas mechanism,  
does not come into effect until 
the energies enter the relativistic domain.
Paradoxically, as the energy rises further
this mechanism becomes less important than the
first-order process due to the contraction of suitable phase space.
In fact it has been shown that the contribution to the
cross section ($\sigma$) given by the Thomas term decreases
 as $\sigma \sim \gamma^{-3}$
(Shakeshaft and Spruch 1979). This is much faster than the
first-order term which behaves as $\sigma \sim \gamma^{-1}$ 
(Shakeshaft 1979)
and ultimately it becomes negligible in the ultrarelativistic region.

In applying the Bates refinement of the {\sc ROBK1}  
approximation (Moiseiwitsch and Stockman 1980),
the results were found to give rather good cross section 
estimates 
for the intermediate energy range. However this model also 
led to the prediction that 
$\sigma \sim \gamma^{-1} (\ln \gamma)^2$ at extremely high
energies. The result that  the
nonorthogonal correction term eventually dominates the primary term 
({\sc ROBK1}) 
was suprising, and subsequently  misgivings have been expressed 
regarding the validity of this approach 
(Eichler 1990). This result also contradicted other
modified first-order models, such as relativistic B1B 
(Moiseiwitsch 1986) which predicted  the result 
$\sigma \sim \gamma^{-1}$ (Toshima and Eichler 1990).
Conversely it could be argued that 
the important influence of this correction term is not more suprising 
than the dominance of second-order terms, and reflects the weakness
of the first-order process.

The picture in second-order theories is not 
so clear.  Most of the work to date has centred on the well-known
{\sc ROBK2} model (Moiseiwitsch 1985, Decker 1990) and 
relativistic impulse approximation {\sc RIA} (Jakubassa-Amundsen
and Amundsen 1980). 
Humphries and Moiseiwitsch (1985) 
followed the standard nonrelativistic approach 
to evaluating 
 {\sc ROBK2} by using a linearised-propagator peaking approximation
as well as making  expansions in $\alpha Z$ 
in the wave functions.  This approximation reduces the $T$-matrix to
a simple analytic form, and gives the result $\sigma \sim  \gamma^{-1}$, 
as in {\sc ROBK1} though with a modified coefficient. 
Decker 
and Eichler (1993) managed to calculate the 
$T$-matrix exactly using numerical techniques, and in doing so
queried the earlier {\sc ROBK2} work on two counts.  Firstly, when 
they compared their results with experimental
data (which involves largish charges) they 
found that {\sc ROBK2} was in worse accord with experiment than 
{\sc ROBK1}.  This discrepancy
has not as yet been fully accounted for 
in that the relative effects of the 
$\alpha Z$ 
expansions and the peaking approximation have not
been isolated.  Decker and Eichler (1993) also claimed signs of
an energy dependence which contained logarithmic terms.
Again possibly
indicative of the size of the corresponding 
coefficient, the  evidence was not conclusive  
even though they performed calculations for $\gamma > 1000$.

Jakubassa-Amundsen and Amundsen (1980) in extending their work on the
impulse approximation to the relativistic regime had argued that
the ionization-like matrix element embedded 
in the {\sc RIA} $T$-matrix would lead to logarithmic
terms. The origin of these terms is the 
retardated potential.
Indeed logarithmic correction
terms are ubiquitous in high-energy atomic processes such 
as the stopping of relativistic charged particles.
In both {\sc ROBK2}
and {\sc RIA} the possibility of retardation 
is obvious from the form of the $T$-matrix but as yet
there has been no detailed mathematical study of both
the origin of these terms as well their 
size in comparison to the {\sc ROBK1} amplitude. We 
have attempted to this effect. We will show  
that the multiplicative factor upon the 
logarithmic terms is of order $\alpha^2( Z_P +Z_T)^2$ compared
with {\sc ROBK1}.

\section{First-order Approximations \label{sect1}}

The cross section for electron capture ($\sigma$), for a given impact 
energy in the laboratory frame,  can be written 
in terms of the scattering amplitude $T(\mbox{\boldmath $ \eta$})$
as follows (Glass \etal 1996):

\begin{equation}
\sigma (\gamma) = (2 \pi \gamma v )^{-2} \int 
 d {\vek  \eta} \ \vert T({\vek \eta}) \vert^2  .
\end{equation}
In this formula, $v$ is the relative speed of target ($T$) and projectile
 ($P$), and $\gamma = (1 -v^2/c^2)^{-1/2}$. The 
momentum transfer variable is denoted by $\mbox{\boldmath $ \eta$}$, 
and all quantities are expressed in atomic units ($
\vert e \vert = m_e = \hbar =1$, and $\alpha = 1/c $). 

Specifically, we consider the capture of an electron 
from the target 1s-state $\Phi_T$ to the projectile 
1s-state ($\Phi_P$). In this paper we use the 
following notation to describe these states:

\begin{equation}
\Phi_{T} = \phi_{T}({\vek r})
\e^{-i E_T t} \, 
\; \; \; \; \; ,  \; \; \; \; \; \; 
\Phi^{\prime}_{P} = \phi^{\prime}_{P}({\vek r^{\prime}})
\e^{-i E_P t^{\prime}}  .
\end{equation}
In the equation above, the functions are defined in the
rest frames of the target and projectile nuclei respectively.
The electron coordinate is denoted by 
$({\vek r})$ with respect to the origin located at the 
target nucleus, and 
it is labelled $({\vek r^{\prime}})$ when referred to 
the projectile nucleus. 
These coordinates are related through the 
usual Lorentz transformation.
All primed variables indicate that they are associated with 
the rest frame of the projectile nucleus. 
The atomic energies, which include the electron rest mass, are
denoted by $E_{P,T}$.
Then the  relativistic first-order OBK (ROBK1) approximation
 takes the form (Shakeshaft 1979):
\begin{equation}
T_{\rm ROBK1}({\vek \eta}) =
 (2\pi)^{3/2}\int d^3{\vek s}\;
 \tilde{\phi}_{P}^{\dag \prime}({\vek K}^{\prime}+{\vek s})
 \tilde{V}_{P}^{\prime}({\vek s})
S \tilde{\phi}_{T}({\vek K})   .
\label{obk1tm}
\end{equation}
The Lorentz transform operator for a boost from the 
target to the projectile frame is denoted by $S$. 
The tilde signifies
the Fourier transform through the definition 
\begin{equation}
 \tilde{f}({\vek p})=
 (2\pi)^{-3/2}\int d^3{\vek r}\;
 \e^{-\i{\vek p}\cdot {\vek r}}\;
 f({\vek r}).
\end{equation}
The momentum transfer vectors are simply given by:
\begin{eqnarray}
{\vek K} =  
{\vek K}_{\perp} + {\vek K}_{\parallel} = 
{\vek \eta} + \hat{\vek v}
 ( \gamma E_T - E_P) / (\gamma v), \nonumber \\
{\vek K}^{\prime} =
{\vek  K}^{\prime}_{\perp} + {\vek K}^{\prime}_{\parallel} = 
{\vek \eta} - \hat{\vek v}
 ( \gamma E_P - E_T) / (\gamma v) . 
\label{mom1}
\end{eqnarray}

Evidently at very high velocities ($\gamma \rightarrow \infty$)
the only energy dependence 
in the expression for $T_{\rm ROBK1}$ is due to the 
operator $S$, and thus $T \sim \gamma^{1/2} $. Then it 
clear that in the same limit $\sigma \sim  \gamma^{-1} $.
This energy dependence is a feature of the
kinematics and is
due to a combination of time-dilation and 
frame transformation.
The cross section for ground-state capture, without change of spin 
($\sigma^{\uparrow \uparrow}_{\rm ROBK1}$) then takes the simple
form, to leading order in $\alpha Z$:
\begin{equation}
\sigma_{\rm ROBK1}^{\uparrow \uparrow}
 (\gamma) \approx { 2^5 \pi Z_P^5 Z_T^5 \over 5 \gamma c^{12}}
\end{equation}
in atomic units.

The difficulty with calculating electron capture using this model 
is that the result, though Lorentz invariant, is not gauge 
invariant due to the nonorthogonality of the intial 
and final states.  Moiseiwitsch and Stockman (1980) investigated 
this problem using a modification 
of the {\sc ROBK1} originally due to Bates.
This defines a correction due to orthogonalization given by:
$T_{\rm DRB} = T_{\rm ROBK1} - T_{\rm corr}$, where, 
\begin{equation}
T_{corr}({\vek \eta}) =
 (2\pi)^{3}\int d^3{\vek s}\;
\tilde{\phi}_{P}^{\dag \prime}({\vek K}^{\prime}+{\vek s})
 S^{-1} \tilde{F}(\bar{\vek s})
S^2 \tilde{V}_{P}^{\prime}({\vek s})
\tilde{\phi}_{T}({\vek K}+\bar{\vek s})  .
\label{tcorr}
\end{equation}
In this expression 
\begin{equation}
 \tilde{F}(\bar{\vek s})
 = (2\pi)^{-3/2}\int d^3{\vek r}\;
 \e^{\i\bar{\vek s}\cdot {\vek r}}
 {\phi}^{\dag}_{T}({\vek r})
 {\phi}_{T}({\vek r})  .
\end{equation}
We use the overline symbol to denote a {\it  stretched} vector 
thus: $\bar{\vek s}={\vek s}_{\perp} + \gamma s_{z} \hat{\vek v}$.
In physical terms this correction term
allows for 
capture through 
elastic scattering followed by the sudden adjustment 
of the wavefunction from target to projectile. 
Although this represents a two-step process,  the
matrix element is  of
first order in the interaction potential.
In mathematical terms the new feature introduced in 
equation (\ref{tcorr}) is given by
 $\tilde{F}(\bar{\vek s})$
which is strongly peaked around $\bar{\vek s}=0$ 
and contains $\gamma$--dependent terms.
The integral can be evaluated by the peaking approximation.
One can show (Glass 1994) that the expression (\ref{tcorr}) 
is strongly determined by this peak, and to a high
degree of accuracy
one can simply replace the momentum  distributions by their
values at this peak:
$\tilde{\phi}_{P}^{\dag \prime}({\vek K}^{\prime}+{\vek s})
\approx \tilde{\phi}_{P}^{\dag \prime}({\vek K}^{\prime})$,
$\tilde{\phi}_{T}({\vek K}+\bar{\vek s}) \approx
\tilde{\phi}_{T}({\vek K})$. Therefore the contribution
of this term has a weight determined by the tails of 
the momentum distributions of the atoms, and as a result will be
of higher order in $\alpha Z$ than the term $T_{\rm ROBK1}$.
After making these assumptions, the integral poses no further difficulty
and we obtain  the following result
for capture without spin flip:
\begin{equation}
T_{\rm corr}({\vek \eta}) \approx
 -{2^5 \pi (Z_{P}Z_{T})^{7/2}
 \left({\gamma +1 \over 2}\right)^{1/2}
 \ln \left[ \gamma (1+v/c)\right]
 \over 
  (v/c) (K^2+Z_{T}^{2})^2
 (K^{\prime 2}+Z_{P}^{2})^2}
\left( 1 + { K^2 \over 2 c^2} \right) .
\label{torr}
\end{equation}
The ($\ln \gamma$) term is a consequence of the
the binary scattering peak 
$\tilde{V}_{P}^{\prime}({\vek s})$ 
and the presence of this factor will ultimately
dominate the energy dependence so that as $\gamma \rightarrow \infty$:
\begin{equation}
\sigma^{\uparrow \uparrow}_{\rm DRB}(\gamma)  \approx 
 \left( { 32 \times 151 \over 21 \times 5 }  \right)
{ \pi Z_P^7 Z_T^7 \over\gamma c^{16}}   
(\ln \gamma )^2
\label{drb}
\end{equation}
in agreement with Moiseiwitsch and Stockman (1980) for capture from
T(1s) to P(1s) without change of spin, and to leading order in 
$\alpha Z_{P,T}$. As mentioned by these authors, such a term 
requires extremely high energies and/or large charges for its
presence to be significant.

\section{Second-order corrections}

The second-order term of the {\sc ROBK} perturbation series gives the
result (Glass \etal 1996):
\begin{eqnarray}
T_{2}({\vek \eta}) =
 \int d^3{\vek s} \int d^3{\vek l}\;
\tilde{\phi}_{P}^{\dag \prime}({\vek K}^{\prime}+{\vek l})
 \tilde{V}_{P}^{\prime}({\vek l})  S^{-1} \nonumber  \\
\times \left[
{ c{\vek \alpha}\cdot ({\vek K}+\bar{\vek l}+{\vek s})
+ \beta c^2+E_T+ \gamma {\vek v} \cdot {\vek l}
\over (E_T+ \gamma {\vek v} \cdot {\vek l})^2 
-c^2 ({\vek K}+\bar{\vek l}+{\vek s})^2-c^4
+\i\epsilon}\right]
S^2 \tilde{V}_{T}({\vek s})
 \tilde{\phi}_{T}({\vek K}+{\vek s}).
\label{t2}
\end{eqnarray}
defined in the limit $\epsilon \rightarrow 0+$.
The Dirac matrices are denoted by ${\vek \alpha}$ and $\beta$.
If a direct comparison is to be made with 
experimental data this 
$T$-matrix should be calculated without 
further approximations.  However, in order to make analytic progress, and 
 if we are primarily concerned with the
asymptotic form of the cross section, we can take $\alpha Z$ to 
be a good perturbative parameter, and seek the leading-order 
term of this integral.
Historically, simplifying assumptions have been used in order to reduce
this integral to a simple analytic form (Bransden and McDowell 1993).
In general these methods relied upon peaking approximations which are 
known to have limited scope. As in section \ref{sect1} 
the basic idea is that if an integrand 
possesses {\it sharp} and {\it isolated} peaks then the integral 
may be simplified.
In expression  (\ref{t2}) 
there are four significant peaks in the integrand at the values:
${\vek l} +{\vek K}^{\prime} = 0, {\vek s} +{\vek K} = 0,$
and ${\vek l}= 0, {\vek s}=0$.
The first pair corresponds to bound-state momentum distributions
while the second pair is associated with binary scattering.  
The fact that the peaks are sharp allows
one to stretch the integration limits from the neighbourhood of the
peak to infinity.
And the  isolation of the peaks permits one to 
separate each term. 
The leading term arises
from the double bound-state peaks 
(${\vek l} +{\vek K}^{\prime} = 0, {\vek s} + {\vek K} = 0$).
However there will be cross terms from the binary and bound-state 
peaks at ( ${\vek l} = 0, {\vek s} +{\vek K} \equiv {\vek t} = 0$ )
and at ( ${\vek l} +{\vek K}^{\prime} \equiv {\vek T} = 0, {\vek s} = 0$).
Although these terms are are of higher order in powers of $\alpha Z$, we
find that that they
have a slower decay as a function of increasing energy and 
therefore 
dominate in the ultrarelativistic limit.
 Finally there will be a term 
of still higher order in $\alpha Z$ that arises from the double
binary peak (${\vek l} = 0, {\vek s} = 0$).
On the understandng  that the peaks are sharp and isolated, 
then the  
expresssion  ({\ref{t2}) can be 
divided into four components as follows:
\begin{equation}
T_{2} = T_{\rm HM}  +T_{\rm BPA} + T_{\rm BPB} + T_{\rm DBP}
\label{4peak}
\end{equation}
The first term ($T_{\rm HM}$) 
was evaluated by Humphries and Moiseiwitsch (1984).
Retaining zeroth and first-order terms
in 
${\vek l}+{\vek K}^{\prime}$
and 
${\vek s}+{\vek K}$
in the propagator,  they obtained the expression:
\begin{equation}
T_{\rm HM}({\vek \eta}) \approx
{ 2^4\pi (Z_{P}Z_{T})^{5/2} \over K^2 K^{\prime 2}}
{ {\vek \sigma}_{P}^{\dag}(-{\bar{\vek K}})
S^{-1}(-c{\vek \alpha}\cdot {\bar{\vek K}}^{\prime}
+ \beta c^2 +\gamma \epsilon_{f})S^2
{\vek \sigma}_{T}({\bar{\vek K}}^{\prime})
\over
(2\gamma E_T E_P -c^4
-\epsilon_{f}^2-K^2c^2+2iZ_{P} \bar{K}c^2
+2iZ_{T} {\bar{K}}^{\prime}c^2)}
\label{fullpk}
\end{equation}
where 
${\vek \sigma}({\vek k})$
denotes a Darwin spinor
\begin{equation}
{\vek \sigma}_{T,P}({\vek k})
= \left( 1+{i {\vek \alpha}\cdot {\vek k} Z_{T,P}
\over 2 k c} \right) w_{T,P}  .
\end{equation}
In this notation the spin vector is defined as 
$ \tilde{w}_{P,T} = (1 \ 0 \ 0 \ 0 )$ for spin up. 
For $\alpha Z_{P,T} \ll 1$  we get
\begin{equation}
T_{HM}({\vek \eta}) =
{ 2^5\pi (Z_{P}Z_{T})^{5/2} 
({\gamma +1 \over 2})^{1/2} F
\over K^2 K^{\prime 2}
(2c^2(\gamma -1)-K^2+2iZ_{P} \bar{K}
+2iZ_{T} {\bar{K}}^{\prime})}
\label{HandM}
\end{equation}
where $F=\gamma +{1 \over 2}\left({\gamma -1 \over
\gamma +1}\right)$ for nonflip ($\uparrow \uparrow$)and 
$F =x(2\gamma +1)\eta /2c$ for capture with spin flip ($\uparrow \downarrow$).

This approximate form for the {\sc ROBK2} transition amplitude has
to be qualified on several counts. Firstly, we 
would expect the peaking approximations used in this derivation to
break down for large charges and/or low velocities; 
neither criterion is strongly satisfied for much of 
the experimental data (Anholt 1985). However, it
transpires that total cross sections within this approximation are
in considerably better agreement with experiment than the {\sc  ROBK1} 
results, which are commonly five times too large. On average the
Humphries and Moiseiwitsch (HM) approximation to {\sc ROBK2} is around a factor
of two lower than {\sc ROBK1}. This still leaves a considerable
gap between theory and experiment. Moiseiwitsch (1988) introduced the
`averaging approximation' in which the external $K^{\prime 2}$ $K^2$
factors in the denominator of equation (\ref{HandM}) are replaced
by $K^{\prime 2}+Z_P^2$ 
and $K^2 +Z_T^2$, which resulted  in a dramatic improvement between
theory and experiment.

In 1990 Decker calculated {\sc ROBK2} without 
approximation and using fully relativistic wavefunctions.
It was discovered that in many cases
rather than improving on the
{\sc ROBK1} predictions, exact {\sc ROBK2} was an order of
magnitude {\it greater } than {\sc ROBK1} (Eichler 1990), and 
consequently in poorer accord with experiment. This obviously
raises serious doubts concerning the validity of the various
peaking approximations discussed above, especially for large
charges and energies in the lower range of the relativistic 
regime. There has been considerable
debate in the literature as to which method is correct, with
Decker and Eichler (1993) arguing in favour of
exact numerical calculation of the perturbation series, while
Moiseiwitsch (1995) has pointed out the possible pitfalls of
simultaneously expanding along the Born series and in 
terms of $\alpha Z$ in the wavefunctions.

\subsection{Binary scattering peaks}

In section \ref{sect1} it was shown that 
the binary scattering
peak can produce logarithmic terms though with a small 
coefficent. The contribution of such terms
within second-order perturbation theory 
can be evaluated without much difficulty so long
as the assumption that the peaks are sharp and isolated
holds true. This supposes, for example, 
 that the peaks at 
 ${\vek l}={\vek 0}$ and
${\vek l}=-{\vek K}^{\prime}$ are distinct, which is certainly
the case so long as
$\alpha Z_P \ll 1$. In this calculation we 
continue to 
make this assumption, using $\alpha Z$ as a perturbation parameter, in
order to obtain the leading-order correction from the binary peaks.

The first binary peak of (\ref{4peak}) leads to the term:
\begin{eqnarray}
T_{\rm BPA}({\vek \eta}) \approx
 \int d^3{\vek t}
 \int d^3{\vek l}\;
\tilde{\phi}_{P}^{\dag \prime}({\vek K}^{\prime}+{\vek l})
 \tilde{V}_{P}^{\prime}({\vek l}) 
 S^{-1} \\ \nonumber
\times
\left[
{ c{\vek \alpha}\cdot \bar{\vek l}
+\beta c^2+E_T+\gamma {\vek v} \cdot {\vek l}
\over 
E_T^2 -c^4+2\gamma E_T 
{\vek v}\cdot {\vek l} -2 c^2 
{\vek t}\cdot \bar{\vek l}-c^2 l^2
+\i\epsilon}\right]
\times S^2 \tilde{V}_{T}(-{\vek K})
 \tilde{\phi}_{T}({\vek t}).
\end{eqnarray}
and a similar term will aries from the second peak at:
 ${\vek l} +{\vek K}^{\prime} \equiv {\vek T} = 0, {\vek s} = 0$.

In the usual manner, we can invert  the Fourier 
transform for $\tilde{\phi}_{T}({\vek t})$ by 
using the identity
\begin{equation}
(\lambda +\i\epsilon)^{-1} = -i \int_{0}^{\infty}du
\exp (i u(\lambda +i \epsilon ))  
\end{equation}
to convert the denominator into an exponential.  
Explicitly, we have for 
1s-states the expression:
\begin{eqnarray}
T_{\rm BPA}({\vek \eta}) \approx
(2 \pi)^{3/2} (Z_{T}^3 / \pi)^{1/2} 
\tilde{\phi}_{P}^{\dag \prime}({\vek K}^{\prime})
\tilde{V}_{T}(-{\vek K})
 \int d^3{\vek l}\;
 \tilde{V}_{P}^{\prime}({\vek l})
  S^{-1}
\nonumber\\
\times \left[
{ c{\vek \alpha}\cdot \bar{\vek l}
+\beta c^2+E_T+\gamma {\vek  v} \cdot {\vek l}
\over
E_T^2 -c^4+2\gamma E_T 
{\vek v}\cdot {\vek l} - c^2 {l}^2
+2 i c^2 Z_{T} \bar{l}+i \epsilon
}\right]
 S^2 
{\vek \sigma}_{T}(-\bar{\vek l})
\end{eqnarray}
To leading order in $\alpha Z$ we have
\begin{equation}
T_{\rm BPA}({\vek \eta}) =
(2 \pi)^{3/2}
( Z_{T}^3 /  \pi )^{1/2} 
\tilde{\phi}_{P}^{\dag \prime}({\vek K}^{\prime})
\tilde{V}_{T}(-{\vek K})
S^{-1} (\beta +1 ) S^2
w_T L
\label{2nd}
\end{equation}
with
\begin{equation}
L =- \int d^3{\vek l}\;
 \tilde{V}_{P}^{\prime}({\vek l})
{1 \over
( {l}^2 -2{\vek v} \cdot \bar{\vek l} 
+Z_{T}^2 -2 i Z_{T} \bar{l}- i \epsilon)}
\label{Lint}
\end{equation}
Even after these simplifications, the calculation of this integral is not 
straightforward. 
However, it is clear that the main feature of the integrand 
is due to the singularity 
in the propagator. It is this factor rather than the potential scattering 
in (\ref{Lint}) that gives rise to the logarithmic behaviour. 
An analytic form can be found for this integral
(Glass 1994) though its form is rather complicated. Instead 
it can be shown to a fair
approximation that this complicated form reduces to,
\begin{equation}
L \approx  - i (2\pi{)}^{3/2}(Z_P /  \gamma v) 
 \ln  ( 2  i \gamma Z_T /  v )  .
\end{equation}

Then within these approximations we have a simple
formula for ground-state to ground-state capture without 
change of spin: 
\begin{equation}
T^{\uparrow \uparrow}_{\rm BPA} \approx - ( 2 i Z_P  / v) 
 \ln  ( 2  i \gamma  v / Z_T)  T^{\uparrow \uparrow}_{\rm ROBK1}. 
\end{equation}
Using the same method we find that 
the term due to the peaks around 
${\vek l} +{\vek K}^{\prime} \equiv {\vek T} = 0, {\vek s} = 0$.
is approximately given by:
\begin{equation}
T^{\uparrow \uparrow}_{\rm BPB} \approx - ( 2 i Z_T  / v) 
 \ln  ( 2  i \gamma  v / Z_P)  T^{\uparrow \uparrow}_{\rm ROBK1}. 
\end{equation}
While these terms are 
of higher order than {\sc ROBK1} in $\alpha Z$, they are 
are of lower order than the Bates term given by 
equation (\ref{torr}). 
Furthermore it is clear that in the limit of extreme relativistic energies
that this term decreases much more slowly than the 
bound-state peaking term (\ref{HandM}). 
If we combine these terms according to (\ref{4peak})  
the result is that
at relativistic energies the nonradiative electron capture cross 
section has the asymptotic form:
\begin{equation}
\sigma^{\uparrow \uparrow}_{\rm ROBK2}(\gamma) \sim  
{2^7 \pi Z_T^5 Z_P^5 (Z_T +Z_P)^2 \over 5 c^{14} }
{(\ln \gamma)^2 \over \gamma}
\end{equation}
It is clear from this expression that the {\sc  ROBK2} 
approximation contains logarithmic terms and that these are in general
more important than the 
term from the Bates approximation (\ref{drb}) for small $\alpha Z$. 
The simple reason is that only one momentum tail 
is sampled in producing the term (\ref{2nd}). To complete our
discussion, we note that additional logarithmic terms will be
produced by the product of the two binary peaks: term $T_{\rm DBP}$ in
equation (\ref{4peak}). However, 
these will be of higher order in $\alpha Z$ than the term (\ref{2nd})
because they include terms from both momentum tails as in 
the Bates approximation. Hence their contribution will 
not be a significant factor in this process for $\alpha Z \ll 1$. Finally 
we conclude by mentioning that capture with spin-flip through
this process contributes  a small additional term.
Further consideration of equation (\ref{2nd}) for the case of
capture with  spin-flip shows
that, in common with {\sc ROBK1}, the
cross section is only a
quarter of the non-flip result ($\sigma^{\uparrow \downarrow}
= {1 \over 4}\sigma^{\uparrow \uparrow}$).
Previously, calculations using 
{\sc ROBK2} (with bound-state peaking) and the
{\sc RCDW} approximation (Glass \etal 1994), showed that 
for  relativistic energies 
capture with spin flip is more probable 
than capture without spin flip.
From the analysis above, we see that this is
not the case in the extreme relativistic regime.

\section{Conclusions}
In the Bates approximation logarithmic effects arise from
scattering at the on-shell binary peaks.
Our estimate of the corresponding terms in
second-order perturbation theory confirm that 
there are indeed logarithmic terms, and the magnitude
of these terms is larger than those obtained from the nonorthogonal
correction introduced by Moiseiwitsch and Stockman (1980).
This supports
the claims made by Decker (1990) and 
Jakubassa-Amundsen and Amundsen (1985) in highlighting the 
cause of this effect and, more importantly estimates the coefficient
of the logarithmic term. 
The total cross section including transitions involving spin flip 
is given by the formula:
\begin{equation}
\sigma_{\rm ROBK2}(\gamma) \sim  
2^5 \pi \alpha^{14} Z_T^5 Z_P^5 (Z_T +Z_P)^2  \gamma^{-1}
(\ln \gamma)^2
\end{equation}
Interestingly,  
the logarithmic correction is {\em not} principally the
result of retardation as previously supposed (Jakubassa-Amundsen and 
Amundsen 1985)
but rather due to propagator singularities.
In other words, it is a principally a kinematic feature.
Furthermore the correction will only be significant for extreme
energies and thus is primarily of academic interest.
Lastly we note  
that capture without change of spin is the
dominant transition at extremely high energies.

\noindent 
\section*{References} 

\begin{description}
\item Anholt R 1985 {\it Phys. Rev. A} {\bf 31} 3579.

\item  Bransden B H and McDowell M R C 1993 
{\it Charge Exchange and the Theory of Ion-Atom Collisions }
(Oxford: Clarendon Press).
\item  Decker F 1990
{\it Phys. Rev. A} {\bf 41} 6552.
\item Decker F and Eichler J 1993 
{\it J. Phys. B: At. Mol. Opt. Phys.} {\bf 26} 2081.
\item Deco G R and Rivarola R D 1987
{\it J.Phys.B:At.Mol.Opt.Phys.} {\bf 20} 3853.
\item Eichler J 1990 {\it Phys. Rep.} {\bf 193} 165.

\item  Eichler J and Meyerhof W E 1995 
{\it Relativistic Atomic Collisions} (New York: Academic Press).
\item[]  Glass J T,  McCann J F  and Crothers   D S F 1994
{\it J.Phys.B:At.Mol.Opt.Phys.} {\bf 27} 3445.
\item \dash 1996
{\it Z.Phys. D} {\bf 36} 119.
\item[]  Glass J T 1994
{\it Ph. D. Thesis, Queen's University Belfast,  unpublished}.
\item Humphries W J  and Moiseiwitsch B L 1984 
{\it J.Phys.B:At.Mol.Phys. } {\bf 17} 2655.

\item Jakubassa-Amundsen D H and Amundsen  P A 1980
{\it Z. Phys. A} {\bf 298} 13.
%  robk2
\item \dash 1985
{\it Phys. Rev. A} {\bf 32} 3106.
\item Moiseiwitsch B L 1986 {\it J.Phys.B:At.Mol.Phys.} 
{\bf 19} 3733.
\item \dash 1988 {\it J.Phys.B:At.Mol.Phys.} 
{\bf 21} 603.
\item \dash 1995 {\it XIX Int. Conf. Physics Elec. At. Collisions 
(Whistler, Canada)} Book of Abstracts, eds. Mitchell J B A \etal p~107.
%  thomas  ?
\item Moiseiwitsch B L and Stockman S G 1979 
{\it J.Phys.B:At.Mol.Phys.} {\bf 12} L695.
%  bates
\item Moiseiwitsch B L and Stockman S G 1980 
{\it J.Phys.B:At.Mol.Phys.} {\bf 13} 4031.
\item  Mukherjee S C, Sural D P, McCann J F and Shimamura I 1992
{\it Comm. At. Mol. Phys.} {\bf 28} 25.

\item Shakeshaft R  1979.
{\it Phys. Rev.}  A  {\bf 20} 779.
\item Shakeshaft R and Spruch L  1979
{\it Rev. Mod. Phys.} {\bf 51} 369.
\item Toshima N and Eichler J 1990 
{\it Phys. Rev.}  A {\bf 40} 125.

\end{description}
\end{document}